\documentclass{PoS}

\title{CP violation in QCD}

\ShortTitle{CP violation in QCD}

\author{\speaker{Michael Creutz}\thanks{
    Notice: This manuscript has been co-authored by employees of
Brookhaven Science Associates, LLC under Contract No. DE-SC0012704
with the U.S. Department of Energy. The publisher by accepting the
manuscript for publication acknowledges that the United States
Government retains a non-exclusive, paid-up, irrevocable, world-wide
license to publish or reproduce the published form of this manuscript,
or allow others to do so, for United States Government purposes.
}\\
        Brookhaven National Laboratory\\
        E-mail: \email{mike@latticeguy.net}}

\abstract{ Among the parameters of QCD is one that results in CP
  violation when non-vanishing.  This is closely related to possible
  quark mass terms.  It is conventionally interpreted in terms of
  gauge field topology or alternatively in terms of phases in the
  quark masses.  There is no experimental evidence for this parameter
  having a non-zero value, a puzzle for theories involving
  unification.  }

\FullConference{XIII Quark Confinement and the Hadron Spectrum - Confinement2018\\
		31 July - 6 August 2018\\
		Maynooth University, Ireland}

\begin{document}


\long \def \blockcomment #1\endcomment{}
\def\slashchar#1{\setbox0=\hbox{$#1$}           
   \dimen0=\wd0                                 
   \setbox1=\hbox{/} \dimen1=\wd1               
   \ifdim\dimen0>\dimen1                        
      \rlap{\hbox to \dimen0{\hfil/\hfil}}      
      #1                                        
   \else                                        
      \rlap{\hbox to \dimen1{\hfil$#1$\hfil}}   
      /                                         
   \fi}                                         %


\abstract{Among the parameters of QCD is one that results in CP violation when
non-vanishing.  This is closely related to possible quark mass terms.
It is conventionally interpreted in terms of gauge field topology or
alternatively in terms of real chiral eigenvalues of the Dirac
operator.  There is no experimental evidence for this parameter having
a non-zero value, a puzzle for theories involving unification.
}

\section{Introduction}

QCD depends on a remarkably small number of parameters.  In
conventional discussions these include the strong coupling constant
$\alpha_s$, the quark masses $m_i$, and the topological parameter
$\Theta$.  But unlike with QED, the theory of electrons and photons,
the connection with physical observables is rather non-trivial.  In
electrodynamics, both the electric charge and the electron mass are
directly observable.

In QCD, because of asymptotic freedom
\cite{Politzer:1973fx,Gross:1973id, Gross:1973ju} and dimensional
transmutation\cite{Coleman:1973jx}, the strong coupling constant is
tied to the overall scale.  That in turn is connected with the
particle masses.  A natural scale to use is the mass of the proton;
once that is determined the strong coupling constant is no longer an
adjustable parameter.  The quark masses are most directly tied to the
pseudo-scalar spectrum.  They can be adjusted to give the correct
masses to the pions, kaons, etc.

The parameter $\Theta$ is perhaps the most subtle.  When non-zero,
this gives rise to CP violation.  It is thus natural to adjust it to
give the correct neutron electric dipole moment.  Since such has not
been seen, we know that $\Theta$ is either zero or very small.  This
is the strong CP problem.

At the heart of these issues is the confinement phenomenon.  The
underlying quarks are not free particles.  The connection to the
scattering of physical particles is subtle, and ambiguities, such as
the so called renormalons, can arise.  These ambiguities are closely
related to defining $\Theta$, a non-trivial task since typical fields
in the path integral are non-differentiable.  The present talk
summarizes some of these issues.  Many of these topics are covered in
considerably more detail in Ref. \cite{Creutz:2018dgh}.

\section {Quark masses and {$\Theta$}}

Looking back a previous editions of this meeting, I see that many of
my contributions were closely related to this subject.  This is
particularly true of my presentation at the 1996 edition in
Como \cite{Creutz:1996wg}.  There I started by considering a naive
variable change on a quark field
\begin{equation}
\psi
\longrightarrow e^{i\gamma_5\Theta} \psi
\label{rotate}
\end{equation}
This will modify the usual mass term
\begin{equation}
\overline\psi \psi \longrightarrow \cos(\Theta)\ \overline\psi \psi
+i\sin(\Theta)\ \overline\psi \gamma_5 \psi
\label{variables}
\end{equation}
This suggests that it might interesting to study
a generalized mass term
\begin{equation}
m\ \overline \psi\psi \rightarrow m_1\ \overline\psi\psi+im_5
\ \overline\psi\gamma_5\psi
\end{equation}
and explore how QCD depends on the two parameters {$m_1$} and
{$m_5$}.  Were the above change of variables valid, one might expect
physics to only depend on the combination $\sqrt{m_1^2+m_5^2}$.  We
will shortly see that this is false.

\begin{figure}
 \centerline{ \includegraphics[height=.25\textheight]{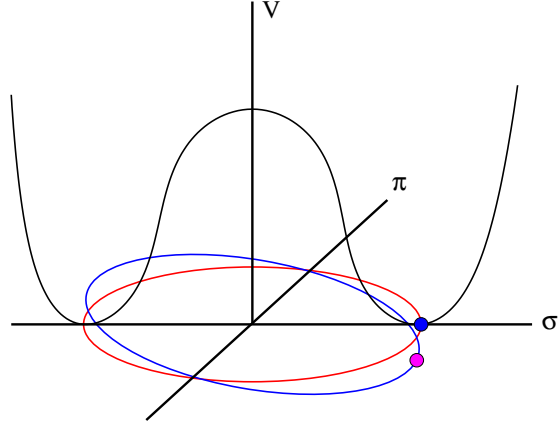}}
  \caption{The starting effective potential.  The mass term results in the minimum being in the positive $\sigma$ direction.}
\label{sombrero}
\end{figure}

Our tool is an effective chiral Lagrangian.  Working with two flavors,
consider the usual ``Mexican hat'' or ``wine bottle bottom'' potential
\begin{equation}
  V=(\sigma^2+\vec\pi^2-v^2)^2 -m_1\sigma.
  \label{effective}
\end{equation}
Here the mass term $m_1\overline\psi\psi \longrightarrow m_1\sigma$
tilts the sombrero, thereby putting the lowest energy state at
positive sigma and giving the pion a small mass $M_\pi^2\propto m_1$.
This is sketched in Fig. (\ref{sombrero}).

So in this picture, what does $m_5$ do?  Indeed, a term like
\begin{equation}
  im_5 \overline \psi\gamma_5\psi \longrightarrow m_5\eta
\end{equation}
does not appear in the above effective potential.  The effect of $m_5$
is of higher order in the chiral theory.  The first thing $m_5$ does
is to induce an expectation value for the eta field
\begin{equation}
  \langle \eta \rangle \propto m_5/M_\eta^2.
 \end{equation} 

To proceed, note that the flavored chiral rotation $\psi \rightarrow
e^{i\tau_3 \gamma_5 \Theta}$ mixes the eta field $i\overline\psi
\gamma_5 \psi\sim \eta$ with the isovector scalar field
$\overline\psi \vec\tau_3 \psi \sim {a_0}_3$.  The
combination $(\eta, \vec a_0)$ represents a chiral pair in direct
analogue with the original combination $(\sigma,\vec \pi).$ 
Flavored chiral symmetry is consistent with a coupling between
these fields of form
\begin{equation}
    \sim \ \left(\pmatrix {\sigma & \vec \pi\cr}\cdot \pmatrix
{\eta\cr \vec a_0}
\right)^2.
  \label{distortion}
\end{equation}
Here the square appears to preserve parity symmetry.

\begin{figure}
 \centerline{ \includegraphics[height=.25\textheight]{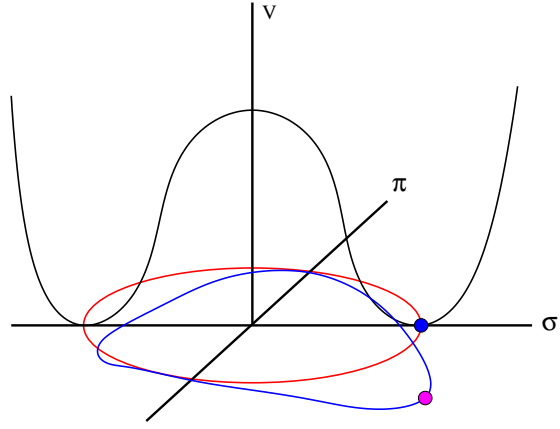}}
 \caption{Including the $m_5$ term in the theory generates a quadratic
   warping of the effective potential.}
\label{m5}
\end{figure}

With an expectation for eta this gives an effective term
\begin{equation}
     (\sigma\eta+\vec\pi \cdot\vec a_0)^2
  \rightarrow \langle\eta\rangle^2 \sigma^2.
  \end{equation}
Including this in our original potential, $m_5$ induces a distortion
proportional to the sigma field squared
\begin{equation}
  V\rightarrow V-\alpha m_5^2 \sigma^2.
\end{equation}
where $\alpha$ is an undetermined constant.  The sign of this term is
related to pi--eta mixing.  This gives rise to a quadratic warping of
the effective potential, as sketched in Fig. \ref{m5}.

\begin{figure}
 \centerline{ \includegraphics[height=.25\textheight]{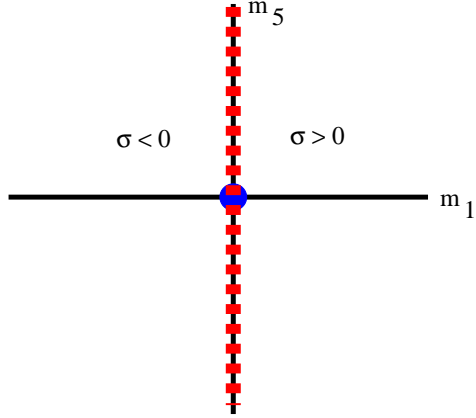}}
 \caption{The phase structure as a function of $m_1$ and $m_5$
   exhibits a first order transition along the $m_5$ axis.  The
   transition is denoted by the dashed line and occurs when the
   conventional parameter $\Theta$ takes the value pi.}
\label{m1m5}
\end{figure}

From this argument, we see that {$m_5$} also gives the pions a mass
\begin{equation}
  M_\pi^2\propto m_5^2.
  \end{equation}
It is crucial to note that unlike the original mass $m_1$ this
is quadratic and not linear in {$m_5$}.  Also note that this term
induces a barrier between {$\sigma>0$} and {$\sigma<0$}.  If we
look at the structure of the theory as a function of the two
parameters $m_1$ and $m_5$, whenever $m_5$ is non-vanishing a first
order transition appears at $m_1=0$.  This is sketched in
Fig. \ref{m1m5}.

The transition occurs when the conventional parameter
$\Theta$ is $\pi$.  The mapping between $\Theta$ and the mass parameters is given by
\begin{equation}
{m_5\over m_1}=\tan(\Theta/2).
  \end{equation}

The crucial conclusion is that although Eq.~(\ref{rotate}) may look
like a harmless change of variables, physics does not only depend on
the combination $\sqrt{m_1^2+m_5^2}$.  The rotation in
Eq.~(\ref{variables}) is ``anomalous,'' and physics depends
non-trivially on $\Theta$.  The important point of this section is
that $m_1$ and $m_5$ are physically independent parameters.

\section{Why is {$\psi \longrightarrow e^{i\gamma_5\Theta} \psi$} not
a symmetry?}

The fact that gamma five rotations are anomalous has been known for
some time \cite{Adler:1969gk,Adler:1969er, Bell:1969ts}.  However a
deep connection between $\Theta$ and the fermionic measure in the path
integral was later elucidated in the work of Fugikawa
\cite{Fujikawa:1979ay}.  When a gauge field configuration has
non-trivial topology, the Dirac operator has chiral zero modes.  The
index theorem relates the number of these modes to the topological
index of the gauge field
\begin{equation}
n_+-n_- =\nu.
  \end{equation}
Here $\nu$ is the gauge field winding number, and $n_+(n_-)$ counts
the number of left (right) zero modes.  Now if we define the trace of
$\gamma_5$ via a sum over the modes of the Dirac operator, we find
that this trace need not vanish
\begin{equation}
{\rm Tr} \gamma_5\equiv \sum_i \langle \psi_i| \gamma_5.
|\psi_i\rangle=\nu
\end{equation}
On configurations carrying topology, the chiral change of variables will introduce a phase in the fermionic measure of the path integral
\begin{equation}
d \psi \rightarrow e^{i \nu \Theta}\ d\psi.
  \end{equation}
Thus this change of variables is equivalent to inserting a factor of $e^{i \nu \Theta}$ into the path integral
\begin{equation}
Z=\int (dA)(d\psi)(d\overline\psi)\ e^{-\beta S}
\longrightarrow
\int (dA)(d\psi)(d\overline\psi)\ e^{i\nu\Theta}\ e^{-\beta S}.
\end{equation}
This represents a physically different theory.  With $\Theta$ present, the theory is CP violating since the $m_5$ term is.
This emphasizes yet further that {$m_1$} and {$m_5$} are inequivalent
parameters that have nothing to do with each other.

A natural question at this point is whether there is an independent
{$m_5$} for each flavor?  The answer is no because flavored chiral
symmetries remain valid.  A change of variables such as
\begin{equation}
\psi \longrightarrow e^{i\gamma_5\lambda_\alpha\Theta} \psi
\end{equation}
is a valid symmetry when $\lambda_\alpha$ is one of the trace less
generators of the flavor group.  Rotations of this form allow us to
move the chiral phase between different flavors.  It is perhaps
amusing to consider rotating any non-trivial $\Theta$ into the top
quark.  In this sense, the top quark does not fully decouple from low
energy physics.

\section{The strong CP problem}

The parameter $m_5$ is inherently CP violating.  But the experimental
bounds on such a symmetry breaking are extremely small, of order
$10^{-10}$ for the ratio $m_5\over m_1$.  Why should this be so?

This is actually only a problem for theories that unify the strong
with the electro-weak interactions.  We know experimentally that the
weak interactions do violate CP.  Thus on reducing energies from the
weak scale, it would be ``natural'' for a residue of this breaking to
to survive in the low energy strong interactions.  On the other hand,
if the strong interactions are never unified with the others, then the
symmetries of the theory make $\Theta=0$ stable under renormalization.

One possible solution to this conundrum involves introducing a new
particle called the axion.  It is arranged to make the parameter
$\Theta$ into a dynamical field with the ground state of this field at
$\Theta=0$.  Effectively this corresponds to adding something like
$(\partial_\mu \Theta)^2$ to the action.  The coefficient of this is
arbitrary, and thus the coupling of the axion particle to the strong
interactions can be made arbitrarily small.

While potentially viable, this procedure seems rather ad hoc and
involves introducing a new particle.  One might also worry that on
descending from the unification scale a linear term in the $\Theta$
field might survive to give an expectation value to the CP violating
effect.

\section{The up-quark mass}

It has been occasionally suggested that a vanishing up-quark mass
would solve the strong CP problem.  The argument is that flavored
rotations can put all phases in the up-quark mass, and if the up-quark
mass vanishes, it cannot have a phase.  The shortcoming of this naive
argument is that the up-quark mass involves the two parameters $m_1$
and $m_5$, and these parameters are independent.  The strong CP
problem is why is $m_5$ small, and has nothing to do with the parameter
$m_1$.

Nevertheless, the possibility of a vanishing up-quark mass has received considerable attention over the years, and thus it is interesting to explore what this means.  To begin with, we know that the pion masses are non-zero, and thus both $m_u$ cannot $m_d$ vanish.  So to proceed we introduce an
up--down mass difference term to the theory
\begin{equation}
m\ \overline \psi\psi \rightarrow m_1\ \overline\psi\psi+m_2
\ \overline\psi\tau_3\psi.
\end{equation}

\begin{figure}
 \centerline{ \includegraphics[height=.25\textheight]{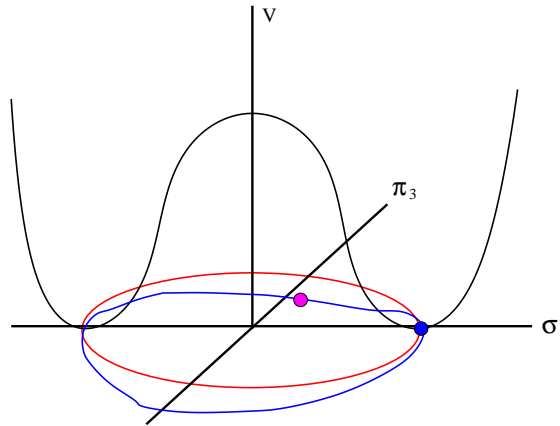}}
 \caption{The presence of a quark mass difference distorts the effective potential downward in the $\pi_3$ direction. }
\label{mdiff}
\end{figure}

The term $m_2\overline\psi\tau_3\psi \sim {a_0}_3$ transforms as an
isovector scalar.  This, just as the $m_5$ term, does not appear
in the starting effective potential of Eq.~(\ref{effective}).  To
study its effect, we parallel the earlier discussion on $m_5$ and note
that $m_2$ will give ${a_0}_3$ an expectation value
\begin{equation}
\langle {a_0}_3 \rangle \propto m_2/M_{a_0}^2
  \end{equation}
This will enter the same effective term as in Eq.~(\ref{distortion})
and warp the effective potential.  But now the distortion is downward
in the $\pi_3$ direction,
\begin{equation}
  V\rightarrow V-\alpha m_2^2 \pi_3^2.
\end{equation}
If we do not include an overall tilt in the $\sigma$ direction, the
{$\pi_3$} field will gain an expectation value!  This represents the so
called the CP violating ``Dashen phase'' \cite{Dashen:1970et}.  This
situation is sketched in Fig. \ref{mdiff}.

\begin{figure}
 \centerline{ \includegraphics[height=.4\textheight]{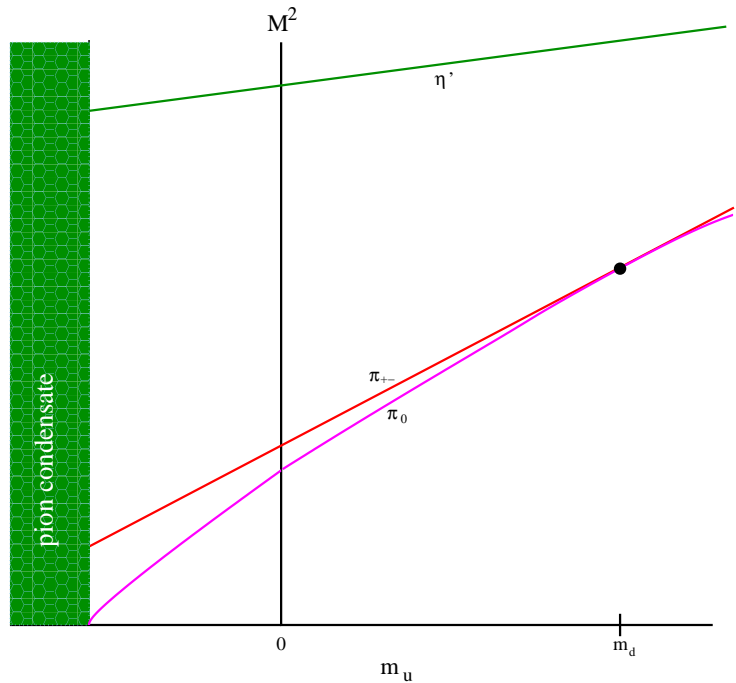}}
 \caption{As the up-quark mass is reduced below that of the down quark, isospin breaking separates the charged and neutral pion.  As the up-quark goes through zero mass, no singularity is expected.  If the up-quark mass becomes sufficiently negative, the neutral pion can condense and we enter the CP violating Dashen phase.}
\label{iso2}
\end{figure}

To understand this phase better, consider fixing the down quark mass
to some positive value and vary the up-quark mass.  As the up-quark
becomes lighter, the pions will decrease in mass, as sketched in
Fig. \ref{iso2}.  Note that the pions remain massive as the up
quark goes to zero.  At that point the mass gap of the theory remains,
and no singularity occurs.  As we continue the up-quark mass into the
negative regime, the pions continue to become lighter, with the
breaking of isospin making the neutral pion lighter than the charged
ones.  But if we make the up-quark mass sufficiently negative, the
neutral pion mass can vanish.  Beyond that point {$\pi_3$} gains its
expectation value, and we are in the Dashen phase.  Note that in this
regime the product of the quark masses is negative and we are formally
at $\Theta=\pi.$

\begin{figure}
 \centerline{ \includegraphics[height=.25\textheight]{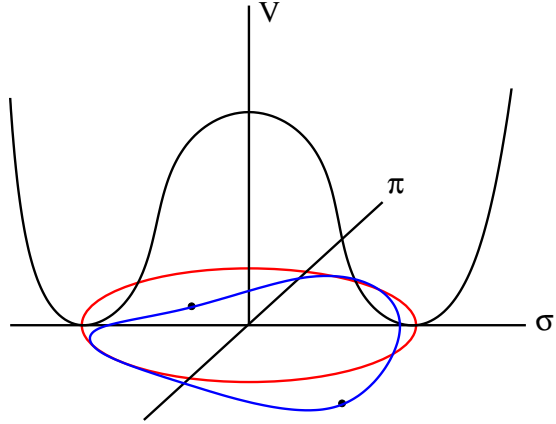}}
 \caption{The four parameters of two flavor QCD map onto the effective chiral Lagrangian as the overall scale, the tilt, the warp, and the angle between the tilt and warp.}
\label{warp}
\end{figure}

We now have at a rather simple picture for the qualitative behavior of
two flavor QCD as a function of the conventional parameters
$\alpha_s\ m_u \ m_d\ \ \Theta$.  These map in a non-linear way into
the overall scale, the tilt of the effective potential, and a possible
quadratic warping.  However in general the tilt and warp need not be
in the same direction, and the fourth parameter represents a possible
angle between them, as sketched in Fig. \ref{warp}.

\begin{figure}
 \centerline{ \includegraphics[height=.3\textheight]{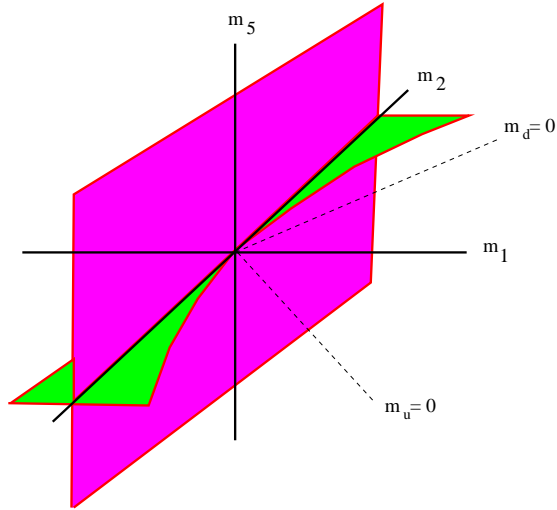}}
 \caption{The full phase diagram for two flavor QCD as a function of
   $m_1,\ m_2,\ m_3$.}
\label{phasediagram}
\end{figure}

Extending these arguments, we can obtain the full two flavor phase diagram
as a function of the parameters $m_1,\ m_2,\ m_3$ \cite{Creutz:2010ts}
as sketched in Fig. \ref{phasediagram}.  The plane at $m_1=0$
extends the first order transition from Fig. \ref{m1m5}.  In addition,
there is another first order transition extending into the $m_5=0$ plane;
the order parameter here is the sign of the non-vanishing expectation
for the $\pi_3$ field.

\section{Symmetries in the masses}

\begin{figure}
 \centerline{ \includegraphics[height=.25\textheight]{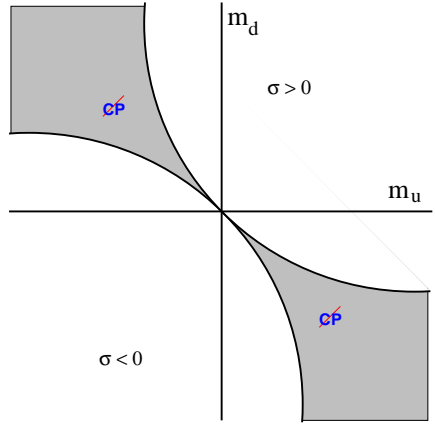}}
 \caption{The phase diagram for two flavor QCD at $m_5=0 $as a function of
   the up and down quark masses.}
\label{iso4}
\end{figure}

This diagram has a variety of symmetries.  First, it is invariant
under changing the sign of the parameter $m_5$.  This is associated
with CP and will protect $m_5$ from any additive renormalization.  What
about the other mass parameters?  To proceed, concentrate on {$m_5=0$}
plane, sketched in Fig. \ref{iso4}.  At the edge of the Dashen phase
is a second order transition where the neutral pion becomes
mass-less.  The order parameter for the transition is the expectation
value for the neutral pion field, $\langle \pi_0 \rangle$.

The next invariance to note is under exchanging the up and down quark
masses.  This is isospin, and protects the quark mass difference $m_2$
from additive renormalization. Then there is a symmetry under
$m_u\leftrightarrow -m_d$.  This represents isospin at $\Theta=\pi$
and protects $m_1$ from additive renormalization.  Another symmetry,
not really independent of the above, is under flipping the signs of
both quark masses.  This can be implemented by a flavored chiral
rotation
\begin{equation}
  \psi \rightarrow e^{i\pi\tau_3\gamma_5} \psi
  \end{equation}
which is not anomalous.

A crucial observation is that this diagram is not symmetric under
$m_u\leftrightarrow -m_u$.  The concept of a vanishing up-quark mass
is not protected by any symmetry!  While symmetries protect the three
parameters $m_1,\ m_2,\ m_3$ individually, these quantities in general
can have independent renormalizations.  One might try to define
\begin{equation}
  \hbox{``}{m_u}\hbox{''}= {m_1+m_2\over 2}+im_5
\end{equation}
however this combines independent parameters and is an artificial
construct.

So this leaves us with a conundrum.  Can any experiment tell if the up
quark mass vanishes?  If not, is $m_u=0$ a well defined concept?  The
issues here are all non-perturbative, so relating the up-quark mass to
a perturbative definition cannot answer this.  Non-perturbative issues
require a regulator such as the lattice.  And at least naively the
lattice can answer the question.  One should adjust the lattice
parameters until the physical hadron spectrum comes out right.  Then
one can read off the input lattice quark masses and see if $m_u=0$.

\begin{figure}
 \centerline{ \includegraphics[height=.15\textheight]{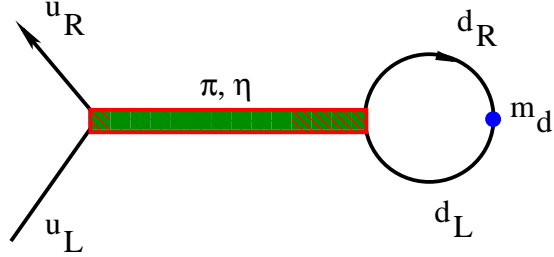}}
 \caption{A non-vanishing down quark mass can, through the pion field,
   convert a left handed up quark into a right handed one.}
 \label{induced2}
\end{figure}

There is a complication to this process.  As shown by 't Hooft
some time ago \cite{'tHooft:1976fv}, a non-vanishing down quark mass
can induce an effective mass for the up-quark.  Both the combinations
$i\overline u \gamma_5 u$ and $i\overline d \gamma_5 d$ couple to the
pion.  This means that through the pion field, a left handed up-quark
can turn into a right handed one by an amount proportional to the down
quark mass.  This is sketched in Fig. \ref{induced2}.  Because of this
effect, ratios of quark masses are not renormalization group invariant
when non-perturbative effects are taken into account.

Can we use the topology of the gauge fields to get a handle on this
issue?  The fermion determinant suppresses non-trivial topology when a
quark mass vanishes.  The concept of a vanishing quark mass is
equivalent to the vanishing of the average gauge field topology.  So
this leaves us with the question of how to define lattice topology.
This is also rather non-trivial.  As is well known, typical
configurations in the path integral involve non-differentiable fields.
Indeed, the space of lattice fields is simply connected in most
formulations.  Topology is lost at the outset since small instantons
can ``fall through the lattice.''  Many studies over the years
\cite{Teper:1985rb,Bruckmann:2009cv} have attempted to get around this
by some sort of cooling process to remove short distance fluctuations
from the gauge fields.  In this way the action is observed to settle
into multiple well defined instantons.  The problem is that while this
procedure is often stable, it is not unique.  The net winding can
depend on the details of the cooling algorithm
\cite{Creutz:2010ec}. Additional questions are what action should we
use to cool, and how long should we cool.

Can we use the index theorem to resolve this? Topology is associated
with zero modes of the Dirac operator.  So we might try to count the
small real eigenvalues of the Wilson fermion operator.  The issue here
is that at finite cutoff these modes are not exact zeros.  How should
we define ``small'' for the eigenvalues?  The result will depend on
the density of real eigenvalues in the first Wilson circle, which in
general need not vanish.  One might instead count zero modes of the
overlap operator \cite{Neuberger:1997fp}, which do occur at the
origin.  The problem here is that the overlap operator is not unique,
depending on a parameter often called the ``domain wall height.''
This parameter is closely related to the Wilson operator and its
eigenvalues.

Should we care if topology is ambiguous?  This is an abstract
theoretical construction and not something directly measured in
laboratory experiments.  One should concentrate on physical
quantities, such as the mass of the eta prime.  The famous
Witten-Veneziano \cite{Witten:1979vv,Veneziano:1979ec} formula does
relate topological susceptibility of the pure gauge theory to the eta
prime mass, but this is only a result in the limit of a large number
of colors.

\section{Summary}

QCD depends on $N_f+1$ possible mass parameters, where $N_f$ is the
number of fermion flavors.  One of these is the CP violating parameter
usually referred to as $\Theta$.  The effects of $\Theta$ are not
visible in perturbation theory.  Indeed, theories with different
values of this parameter have identical perturbative expansions.

Existing experiments have found no evidence for a non-vanishing value
for $\Theta$.  This is a puzzle for models of unification since CP
violation is evident in weak interaction processes.  One possible
solution involves introducing an axion to make $\Theta$ into a
dynamical field that then relaxes to zero.  The possibility of $m_u=0$
is not a viable solution since it involves an unnatural fine tuning.
Three unrelated parameters, $m_1$, $m_2$, and $m_5$, contribute to the
up-quark mass.  Only one of these is associated with CP violation.
The other two have nothing to do with the puzzle.

\end{document}